\documentclass[aps,pre,twocolumn,amsmath,amssymb,amsfonts]{revtex4}
\usepackage{epsfig}
\usepackage{graphicx}
\usepackage{dcolumn}
\usepackage{bm}

\begin{document}

\author{Alexander Jonathan Vidgop$^1$}
\author{Itzhak Fouxon$^{2}$}

\affiliation{$^1$ Am haZikaron Institute, Tel Aviv 64951, Israel}
\affiliation{$^2$ Raymond and Beverly Sackler School of Physics and Astronomy,
Tel-Aviv University, Tel-Aviv 69978, Israel}

\title{Turbulence -  "motion of multitude": a multi-agent spin model for complex flows}

\begin{abstract}

We propose a new paradigm for emergence of macroscopic flows. The latter are considered as a collective phenomenon created by
many agents that exchange abstract information. The information exchange causes agents to change their relative positions which
results in a flow. This paradigm, aimed at the study of the nature of turbulence, appeals to the original meaning of the
word: "turbulence" siginifies "disordered motions of crowds". We give a preliminary discussion on a model of multi-agent dynamics
that realizes the paradigm. This dynamics is reminiscent of spin glasses and neural networks. The model is relational, i. e. it
assumes no spatio-temporal background and may serve as a basis for an approach to quantum gravity via spin models.

\end{abstract}

\maketitle

Turbulence is often said to be the last major unsolved problem of the classical physics. The main obstacle to the research is that
though the flow field is classical, it is effectively random. As a result, the complexity of the phenomenon is the one of a strongly non-linear quantum field theory \cite{Frisch,FGV}. While posing difficulties characteristic of fundamental physics, turbulence is always in our immediate everyday vicinity. Majority of the world around us is fluid, while major part of the fluid flows are turbulent. Yet, despite this ubiquity of the latter flows, that makes them subjects of intense study for centuries, our understanding of turbulence is insufficient. Apparently a certain kind
of order is present in turbulence. In particular, this order underlies certain universal features of the flow statistics that are independent of the stirring force \cite{Frisch,FGV}. It is sometimes surmised that the order is connected with some non-random coherent structures present in the flow. Yet strong non-linearity of the underlying equations hinders the analysis. In this situation it may be fruitful to consider what the turbulent flow constitutes in principle.

Hydrodynamics, in general, and turbulence, in particular, are emergent phenomenona. Macroscopic flows emerge on the basis of complex underlying
events. The latter provide a kind of flow decomposition into "elementary excitations" that could be said to provide a "language of turbulence".
Within the traditional approach the flow arises on the basis of the microscopic molecular dynamics. Here the flow is decomposed into the sum of velocities of particles that move according to the laws of mechanics. This decomposition does not provide a tractable condensation of information on complex flows. However, one may consider alternative decompositions, looking for the one that would give the most efficient way of thinking. Due to the universality of the macroscopic flow description, diversity of such decompositions is unbounded.
Probably the most used alternative decompositions are provided by the lattice-gas automata and the multi-fractal model. Within the lattice-gas automata approach the flow is generated by motion of particles on a lattice according to a certain algorithm \cite{Frisch1}. Within the multi-fractal model the flow
velocity is decomposed into contributions of different fractal sets\cite{Frisch}. However this model is phenomenological and static and it cannot be considered a decomposition into elementary excitations. Another example is provided by the general relativity. It was recently shown that a non-linear fluid flow is contained in a particular type of solutions of the general relativity without matter \cite{Minwalla,EFO}. Thus any underlying microscopic picture of the space-time, for example a spin foam, would give a possible decomposition for a macroscopic fluid flow.
As of today, the search for a tractable language of thinking of complex turbulent flows is open. 

The above questions involve a crucial issue in the search for order of turbulence. Consider the possibility that the order of turbulence is the 
one of very complex discrete events. After continuum approximation is introduced, because of the involved smearing, this order may become hidden. 
In fact, the loss of information occurring in the smearing may in practice prevent one from distinguishing the order. Hence consideration of different bases for the emergence of turbulence may eventually contain the key to the problem. 
Here we consider a paradigm for studies of turbulence that appeals to the primary notions of the phenomenon.

Turba is a Latin word for crowd which is associated with an ancient Greek word meaning disorder \cite{Crowds}. For Romans the word denoted the disorderly mass of people. The derived word turbulence is associated with "disordered motions of crowds" or "disordered motions of multitudes". However, multitudes may be expected to be not quite disordered and have their own complex logic. This would be the logic of information exchange between individual "agents" that makes them change their relative positions, producing a resultant flow. In this way information exchange ultimately leads to a flow. One can ask how complex this flow can be, depending on the mechanism of the information exchange.
It seems plausible that
by choosing a sufficiently complex mechanism one can achieve an arbitrarily complex flow, in particular, a turbulent one. Thus, as part of the search for order in turbulence, it appears reasonable to study the flow formed by many individual "agents" that exchange information. It should be noted that information considered here is of abstract type and not necessarily classical or quantum one. Below "agents" will be called "creits" in accord with the terminology for informational approach to physics established in \cite{VF}.

We now describe a model realizing the above paradigm for emergence of a macroscopic flow. One studies a system of $N$ creits that are different.
It is assumed that the creits can exchange information of $d$ different types or "tones". The paradigm realization can be divided into two parts. In
the first part one introduces appropriate variables that describe the information carried by the creits and determines the law of information exchange. In the second part one determines how the information exchange results in relative motions of the creits and how these motions determine a flow. We pass to address the first part.

Let us list the principles of information exchange assumed in the model. It is assumed that each two creits are potentially able to exchange information (ramifications are discussed below). This exchange is characterized by a certain intensity. This intensity will also be called the
strength of informational connection of the two agents. An important assumption is that the agents are able to "tune" to each other, that is to enter into conditions where the information exchange between them is optimal (to enter into "hed" \cite{VF}). To model this a creit is characterized by a certain directionality or orientation in the information space. This directionality allows the creit to exchange information better with some creits and worse with others. This also allows to incorporate into the model the principle that current informational connections of the creit make it more tuned to information of the same type, while making other types of information "unnoticed" (the law of the "dominant" \cite{VF}).

An effective modeling of the above principles is realized by associating a $d-$dimensional vector ${\bm S}_i$ with each creit, where $i$ is
the creit index. We assume that ${\bm S}$, as a characteristics of informational directionality, has a fixed norm.
Then, the intensity $I_{ij}$ of the information exchange between the creits $i$ and $j$, or the strength of their connection, is
assumed to be given by
\begin{eqnarray}&&
I_{ij}=J_{ij}\left[{\bm S}_i\cdot {\bm S}_j+const\right].
\end{eqnarray}
Here $J_{ij}$ is a constant symmetric matrix of positive coefficients, $J_{ij}=J_{ji}$. The constant is chosen so that $[{\bm S}_i+{\bm S}_j]^2/2=\left[{\bm S}_i\cdot {\bm S}_j+const\right]$ and $I_{ij}$ is always non-negative in accord with its meaning of intensity.
Thus each creit is characterized by its direction in the information space and its abilities $J_{ij}$ for information exchange with other creits. All differences between the creits are encoded in the coefficients $J_{ij}$. We note that it can be expected that in many situations  $i-$th creit can be characterized by $d$ numbers $\xi^a_i$ that characterize its abilities for information exchange with other agents, so that $J_{ij}$ is some function of these numbers $J_{ij}=F(\bm \xi_i, \bm \xi_j)$.

We will assume that there is a finite range of interactions in the information space. Besides qualitative reasons for that, this is needed in order to have a well-defined hydrodynamic velocity field as a result of the modeling. The latter demands finite range of correlations in space. The finiteness allows to define velocity field by spatial averaging over
a scale much smaller than the scale of macroscopic variations, but much larger than the scale of microscopic correlations. Such separation of scales is necessary for a hydrodynamic field to be well-defined, see e.g. \cite{Landau9}. Finite range of correlations in the information space is realized by introducing an informational threshold $I_0$ such that if $I_{ij}<I_0$ then there is no exchange of information between the creits $i$ and $j$. Such threshold is reminiscent of neural networks. Thus the system represents a network where each creit is connected with a finite number of other creits.  One can expect that the most interesting situation arises when this number is large (say for networks appearing in brain modeling one has the total of about $10^9$ neurons such that each neuron is connected to about $10^4$ other neurons, see e. g. \cite{SpinGlass}).

We observe that the expressions for connection strengths are similar to those occurring in studies of spin glasses and neural networks \cite{SpinGlass}. This similarity is further enhanced by the dynamics of spins that we introduce now. We assume that the system evolves naturally to the state where information is exchanged in the most intensive way possible.
Otherwise said we assume the system evolves to the state of maximal connectivity $\sum \theta\left[I_{ij}-I_0\right]I_{ij}$. Introducing the "disconfort"
function
\begin{eqnarray}&&
H\equiv -\sum_{i>j}\theta\left[I_{ij}-I_0\right]I_{ij} ,
\end{eqnarray}
where $\theta(x)$ is the step function, we write the simplest possible dynamics
\begin{eqnarray}&&
\frac{d\bm S_i}{dt}=-\frac{\partial H}{\partial {\bm S_i}}.
\end{eqnarray}
This dynamics is of course quite the same as the one occurring in the theory of the spin glasses \cite{SpinGlass}. The main difference is the finite range of correlations described by the step function. It seems this dynamics is the most natural one can write.

While the above equation can potentially describe a decaying turbulence to have steady turbulence we must add a source to the equation and
consider
\begin{eqnarray}&&
\frac{d\bm S_i}{dt}=-\frac{\partial H}{\partial {\bm S_i}}+\bm f_i.
\end{eqnarray}
Here the (random) source $\bm f_i$ must correlate different spins over a range much larger than the range of their interactions in order to mimic the large-scale
forcing typical for turbulence. Since one may expect to have characteristic domains such that the creits are connected within the domain, the force
must introduce correlations between different domains.

To see if the resulting dynamics produces a flow that is described by equations of fluid
dynamics, one should address the second part of the paradigm - how spin dynamics produces changing distances between the creits. Note that, within our setting, these distances need not necessarily correspond to any "real" spatial distances. The distances must only be some positive numbers assigned to pair of creits and obeying the appropriate geometric limitations such as the triangle inequality. Our ultimate purpose here is only to have a model producing a flow described by the usual hydrodynamic equations.

Thus we have the problem of what can be called "relational physics" \cite{Dreyer1}. We have $N$ objects with a given set of relations between them. The symmetric matrix of relations $R$ is determined by the strengths of connections, $R_{ij}=I_{ij}$ (the choice of the values of $R_{ii}$ is irrelevant for our discussion here). A relation is provided by $I_{ij}$ independently of whether $I_{ij}$
is below or above the threshold $I_0$. We look for a meaningful way
to assign the matrix of distances between the objects $D$ provided the matrix of relations $R$ is given. As the simplest prescription one could try to base the distance between two objects only on the relation between these two objects, and not on the whole set of relations, that is consider a relation of the type $D_{ij}=H(R_{ij})$ where $H$ is some function. Making the assumption that objects which connection is stronger are also closer in space, one considers decreasing positive functions $H$. Then the objects which are far apart will have the connection strength below the threshold and will not interact. This would guarantee a finite range of interactions in space necessary for validity of hydrodynamics. As a concrete prescription one may assume that the distance is given by $(R_{ij})^{-1}$. The problem with this straightforward procedure is that the triangle inequality is generally not satisfied and thus these distances cannot be imbedded in a three-dimensional Eucledian space. Apparently any relation of the type $D_{ij}=H(R_{ij})$ will suffer from this difficulty and one must consider relations where $D_{ij}$ depends on the whole matrix of relations (note that this also seems a more realistic possibility in many situations). Relations such as $D_{ij}=(R^{-1})_{ij}$ are also not appropriate because too many independent degrees of freedom will appear in $D$. Generally speaking, a spatial configuration of $N$ objects is determined by $3N-6$ independent numbers. Here $3N$ corresponds to the total number of coordinates of $N$ objects while the subtraction of $6$ gets rid of shift and rotations. On the other hand, the number of distances is $N(N-1)/2$ so these must satisfy certain constraints to represent a configuration of objects in ordinary space. In other words the matrix $D$ must not depend on all $R_{ij}$ separately.

The problem of recovering the notion of the distance from pure relations between things is connected with the theory of the quantum gravity. It is widely believed that in constructing this theory, eventually, one will have to return to the viewpoint of Leibniz and derive physics from a purely relational basis, see e.g. \cite{Dreyer1}. According to this viewpoint, matter exists as a complex network of inter-relationships between its components, while the space-time "container" is an abstraction, a framework that is put on this network externally. On the most fundamental level, this framework is not able to "hold" things and pure relations become fundamental. In fact, the (quantum) Heisenberg spin model, that our model closely resembles, was considered as a model for the relational basis out of which the spatio-temporal metric would emerge \cite{Dreyer1}. Our modeling suggests a system of many agents exchanging information as the origin of the fundamental relational spin model.

While the derivation of the correspondence between the set of relations and the spatial configuration of creits is beyond the scope of the present work, let us make a comment on the feasibility of flows modeling on the base of the spin models. As indicated in \cite{Dreyer1} and especially in \cite{Wen},  spin models can produce extremely diverse emergent behavior. Besides the space-time metric, they can give rise to fermions and gauge interactions in the low energy limit. It is the reasonable to expect that spin models are able to reproduce hydrodynamics. Let us mention in this context that
because our model is a relational one, it should lead to no difficulties with Galilean invariance, in contrast to models based on fixed lattice such as lattice-gas automata.

We have presented a new paradigm for the study of emergence of macroscopic flows and turbulence. The flows are suggested to emerge on the
basis of exchange of information between agents that makes them change their relative positions. Due to the basic nature of the setting, the paradigm can be expected to find many applications. We established the foundations of a spin model giving concrete realization to the paradigm. This model is reminiscent of the dynamics of spin glasses with finite range of correlations and it is interesting in its own right. When the force acting on the spins is a large-scale one and it correlates different domains, one can expect turbulent regimes to occur. The model, though a classical one,
allows natural quantization. It describes evolution of relations between agents and it can be considered as a basis for an approach to the problem of quantum gravity. The problem of establishing the correspondence between the relational network and a spatial configuration of objects is posed.

%%%%%%%%%%%%%%%%%%%%%%%%%%%%%%%%%%%%%%%%%%%%%

%%%%%%%%%%%%%%%%%%%%%%%%%%%%%%%%%%%%%%


\begin{thebibliography}{99}
%%%%%%%%%%%%%%%%%%%%%%%%%%%%%%%%%%%%%%%%%%%%%

\bibitem{Frisch} U. Frisch, \textit{Turbulence: The Legacy of A. N.
Kolmogorov} (Cambridge University Press  1995).

\bibitem{FGV} G. Falkovich, K. Gawedzki, M. Vergassola
%Particles and fields in fluid turbulence,
Rev. Mod. Phys. {\bf73} 913-975 (2001).

\bibitem{Frisch1} U. Frisch, B. Hasslacher, and Y. Pomeau, Phys. Rev. Lett. \textbf{56}, 1505 (1986).

\bibitem{Minwalla}
  S.~Bhattacharyya, V.~E.~Hubeny, S.~Minwalla and M.~Rangamani,
  %``Nonlinear Fluid Dynamics from Gravity,''
  JHEP {\bf 0802}, 045 (2008).

\bibitem{EFO} C. Eling, I. Fouxon and Y. Oz, Phys. Lett. B {\bf 680},  496 (2009).

\bibitem{Crowds} J. T. Schnapp and M. Tiews, Eds., \textit{Crowds} (Stanford University Press, Stanford 2006).

\bibitem{VF} A. J. Vidgop and I. Fouxon, arXiv:0907.0471.

\bibitem{Landau9} E. M. Lifshitz and L. P. Pitaevskii, \textit{Physical Kinetics}  (Butterworth-Heinemann, Oxford 1981).

\bibitem{SpinGlass} M. Mezard, G. Parisi, and M. A. Virasoro, \textit{Spin Glass Theory and Beyond} (World Scientific, 1987).

\bibitem{Dreyer1} O. Dreyer, arXiv:gr-qc/0404054.

\bibitem{Wen} X-G. Wen, \textit{Quantum Field Theory of Many-Body Systems} (Oxford University Press, Oxford 2004).

%%%%%%%%%%%%%%%%%%%%%%%%%%%%%%%%%%%%%%
\end{thebibliography}
\end{document}